\documentclass[draft]{agujournal2018}
\usepackage{apacite}
\usepackage{url} 
\usepackage{lineno}
\draftfalse

\begin{document}
	
	%
	%
	
	
	\title{Changes in long-term properties of the Danube river level and flow induced by damming}
	
	%
	%
	
	
	
	
	\authors{Djordje Stratimirovi\'{c}\affil{1}, Ilija Batas-Bjeli\'{c}\affil{2}, Vladimir Djurdjevi\'{c}\affil{3}, and Suzana Blesi\'{c}\affil{4}}

	
	\affiliation{1}{University of Belgrade, Faculty of Dental Medicine, Dr Suboti\'ca 8, 11000 Belgrade, Serbia}
	\affiliation{2}{Serbian Academy of Sciences and Arts, Institute of Technical Sciences, Knez Mihailova 35, Belgrade, Serbia}
	\affiliation{3}{University of Belgrade, Faculty of Physics, Institute for Meteorology, Studentski trg 16, Belgrade, Serbia}
	\affiliation{4}{Ca'Foscari University of Venice, Department of Environmental Sciences, Informatics and Statistics, Science Campus, Via Torino 155, Mestre-Venice, Italy}
	
	
	
	
	\correspondingauthor{Suzana Blesi\'{c}}{suzana.blesic@unive.it}
	
	
	
	
	\begin{keypoints}
		\item Changes in long-term Danube river dynamics, caused by Djerdap/Iron Gates dams, are presented;
		\item Our results point to the alterations in natural cycles in river level time series;
		\item Our results point to a complete or partial loss of the annual cycle upstream from dams.
	\end{keypoints}
	
	%
	%
	
	
	\begin{abstract}
		In this paper we assessed changes in scaling properties of the river Danube level and flow data, associated with building of Djerdap/Iron Gates hydrological power plants positioned on the border of Romania and Serbia. We used detrended fluctuation analysis (DFA), wavelet transform spectral analysis (WTS) and wavelet-based modulus maxima method (WTMM) to investigate time series of river levels and river flows recorded at hydrological stations in the vicinity of dams and in the area of up to 480 km upstream from dams, and time series of simulated NOAA-CIRES 20th Century Global Reanalysis precipitation records for the Djerdap/Iron Gates region. By comparing river dynamics during the periods before and after construction of dams, we were able to register changes in scaling that are different for recordings from upstream and from downstream (from dams) areas. We found that damming caused appearance of human-made or enhancement of natural cycles in the small time scales region, which largely influenced the change in temporal scaling in downstream recording stations. We additionally found disappearance or decline in the amplitude of large-time-scale cycles as a result of damming, that changed the dynamics of upstream data. The most prominent finding of our paper is a demonstration of a complete or partial loss of annual cycles in the upstream stations' data that we found to extend as far as 220 km from dams. We discussed probable sources of such found changes in scaling, aiming to provide explanations that could be of use in future environmental assessments.
		
	\end{abstract}

	%
	%
	
	%
	
	
	%
	%
	%
	%

	\section{Introduction}
	
	Since the initiation of present-day scaling techniques in statistical hydrology \citep{Hurst1951}, the role of stochasticity in climate, and specifically stochasticity in river flow dynamics, has been extensively studied. The original work of Hurst \citep{Hurst1951, Hurst1956} demonstrated the existence of a power-law-type time dependence of statistical functions describing river discharges that seamed to be widespread in the dataset of world rivers available at the time. This was an empirical proof that "the natural phenomena [so far considered] have a similarity amongst themselves but differ from purely chance phenomena" \citep{Hurst1956}, later interpreted as a symptom of a long-range dependence, or of an infinite internal memory \citep{Mandelbrot2001}. Statistics and dynamics of river discharges have been since analysed in a number of studies that used both traditional statistical methods, and methods inspired by the analyses of Hurst. A plethora of thus produced further empirical results confirmed the original findings of the existence of scaling in time series of river levels and flows, informing on the ubiquity of memory in river flow dynamics (see, e.g., \citet{Kantelhardt2003a} and \citet{Bunde2013a} and references therein, \citet{Koutsoyiannis2005} and studies of scaling in river flows mentioned therein, or \citet{Mandelbrot1968, Mandelbrot1969}, \citet{Livina2003}, \citet{Bogachev2012} and \citet{Bunde2013}), and extending them to the research of details of complexity of this dynamics, including comprehensive examination of nature and sources of non-stationarity in river hydrological records \citep{Klemes1974, Klemes1987, klemevs1997water, Vanmarcke1983, Mesa1993}, or multifractality of its behavior \citep{Kantelhardt2003a}. We offer the addition to this body of knowledge that examines human-induced alterations in long-range order of river flows caused by damming, aiming to provide research from which future interdisciplinary programs linking hydrology and hydro power with climate and climate change can arise \citep{Lundberg2017a}.
	
	According to the World Commission on Dams \citep{WCD2000} "less than 2.5\% of our water is fresh, less than 33\% of fresh water is fluid, less than 1.7\% of fluid water runs in streams", while human interventions have been stopping even this, for "we dammed half our world’s rivers at unprecedented rates of one per hour, and at unprecedented scales of over 45 000 dams more than four storeys high". As an important technical innovation of humankind, dams are supporting our living by regulating river flows for flood control, irrigation support and electricity production, and continue to hold central stage in recent years of increasing desire for non-fossil fuel-based energy \citep{WCD2000, Lundberg2017a, Klemes2002, KoutsoyiannisD.2017}. At the same time, however, researchers in ecohydrology, closely following these developments, provide evidence about ecological consequences of hydro power river flow management and regulation, such as the decrease in water quality and impact on the exchange of sediments, nutrients, and organisms between and among aquatic and terrestrial regions \citep{Teodoru2005, Klaver2007, Pavlovic2016}, or causal role in species shifts and increased mortality rates of aquatic species migrating downstream \citep{Bacalbasa-Dobrovici1997, Brezeanu2006, Martinovic-Vitanovic2013}. Damming is additionally reported to induce an important impact on changes in biogeochemical river cycles \citep{Friedl2002}, resulting, among other, in methane emissions contributing to global climate change \citep{WCD2000, Lundberg2017a}. With this in mind, in this study we specifically investigate how damming affects scaling dynamics and cyclical consistency, in the case study of the Danube river flow. 
	
	Danube is the second largest river in Europe, with a total length of 2875 km and a catchment area of 817000 km$^2$ \citep{Vukovic2014}. Its importance spans a vast variety of research interests, from studies of the water balance along the river, in the delta \citep{Poncos2013, Mierla2015}, and of the Black Sea (where the Danube contributes with over 60\% of the total inflow into the sea) \citep{Levashova2004}, to researches of the anthropogenic pressure that the 10 countries and over 165 million persons that it connects exert on its dynamics, continuum \citep{Vannote1980}, and surroundings \citep{bloesch2006ultimate, Martinovic-Vitanovic2013, costlow2017water}. Modern hydraulic interventions in the Danube river basin resulted in the construction of eight dams in the period 1956–-1985; of those the most ambitious waterworks, the hydro power and navigation systems of Djerdap (or Iron Gates) I and II, were constructed over the period 1964–-1985 by the joint efforts of Romania and former Yugoslavia (SFRY). The dam for the hydro power plant Djerdap (or Iron Gate) I was constructed in 1972, positioned at 943 km from the Black Sea, producing a formation of the reservoir of volume of 3500 million m$^3$ under average hydrological conditions \citep{Vukovic2014}, an increase of about 2100 million m$^3$ compared to the previous (natural) channel. By 1984, the second dam, Djerdap (or Iron Gate) II was constructed and operational, 80 km downstream from the first dam, built basically to compensate the regulation of water level in the lower pool of the first dam \citep{Levashova2004}. The construction of these dams was thus a big enough endeavour to substantially change the morphology of the river and disturb its natural equilibrium both upstream and downstream. Having in mind that the two dams were constructed in the relative vicinity to each other, and that the whole intervention is performed inside a canyon that river Danube forms in this region, we hypothesized that this would present with a conveniently confined natural system that will allow to study the influence of damming on the river flow dynamics. It was our presumption that, provided that reasonably long historical records are available, and in the absence of significant effects of any other major local hydroclimatic mechanism \citep{Lima2017}, any changes in the river dynamics that we find when comparing records in the periods before and after constructions of dams can be attributed to the change in physical conditions at the location of individual hydrological measuring stations that is induced by damming. 
	
	We present the geographical map of the catchment area around dams in Figure \ref{Fig1}. According to K\"{o}ppen-Geiger climate classification \citep{Kottek2006}, great part of the catchment upstream from dams is a warm temperate-filly humid (Cf) type, with dominating Cfb sub-type (warm summers) and with smaller area covering Cfa sub-type (hot summers). On the borders of catchment, in mountains areas, boreal-full humid with warm (Dfb) and cool summers (Dfc) type is present. Just on the western margins of catchment, in the Eastern Alps, ET type (polar tundra) can be found in some locations. In the geographic area surrounding the dams, for dominating climate types, precipitation is present throughout all months in the year. Generally, in the colder part of the year, precipitation is dominantly linked to extratropical cyclones, and in the warmer part of the year precipitation is linked with convective thunderstorm systems. During the winter snow substantially contributes to precipitation totals. Monthly precipitation maximum is observed for the months at the beginning of summer, but also secondary maximum can present in the beginning of winter, with often substantial contribution of snow.
	
	Upon the construction of power plants Djerdap/Iron Gates I and II, there has been significant human-made contributions to the changes of river levels and flows. The operation of these hydro power plants was a jewel of the SFRY energy planning, due to its significance and shared international interests with neighbouring Romania. On SFRY side, the operation of dams was discussed with Romanian planers before the commission of hydro power plants, and it was decided to use the river level at river Nera confluence for all arbitrages \citep{Jakovljevic1979}. SFRY energy planning was performed as hydro-thermal coordination, based on the fruitful scientific cooperation with Sweden. First, the levels of weekly storage of dams were optimally designed a year in advance, according the previous 50 years of daily measurements. Later, fine tuning of planning was done on a weekly scale, using the methodology of variable and constant energy. Finally, for Djerdap/Iron Gate I (DAM1), which operates as storage dam (the production of electricity may be postponed for around a day), the hourly operation was also defined one day in advance. Typically, this daily chronological production diagram compromises two peaks, daily peak at noon and evening peak at 19.30. The operation of Djerdap/Iron Gate II (DAM2) is similar, but it operates as a run-of-river dam; for regulation of dams nationally developed computer tools PROCOST and ASTRA are used. Finally, half of the dams operational system is regulated separately by Romanian planners, with real time monitoring and yearly arbitrage according to the common dispatching protocol \citep{Jakovljevic1979}. 
	
	We approach our research hypothesis by using the 2nd order detrended fluctuation analysis (DFA2) to describe scaling dynamics, or long-range autocorrelations of the Danube river flow by determining the DFA or Hurst scaling exponent $\alpha$. We use DFA2 in combination with the wavelet transform (WT) power spectral analysis, to confirm DFA2 results (by determining the scaling exponent $\beta$ of the wavelet power spectra), and to additionally examine cycles and cyclical consistency of our records. Previous studies of the long-range dependence, or long-term persistence (LTP) of the Danube river flow report on the existence of LTP in the Danube flow records, with the Hurst exponent values of $\alpha$ in the range from $\alpha=0.67$ to $\alpha=0.85$ for measuring stations in Germany, Austria, Slovakia, Hungary, and Romania, and a crossover in scaling behaviour at time scales $n_c \approx 2-30$ days (or, alternatively, at 20--100 days) \citep{Janosi1999, Kiraly2002, Kantelhardt2003a, Koscielny-Bunde2006, Kantelhardt2006, Bunde2013a, Szolgayova2014}. In addition to these linear long-range autocorrelations, some of these researches inform on the distinct non-linear long-range character of the Danube river flow, manifested in strong multifractality of its records \citep{Kantelhardt2003a, Koscielny-Bunde2006, Kantelhardt2006, Bunde2013a}. In this, dynamical sense, Danube is not very dissimilar from other world rivers; based on these and similar findings, a general assumption was made that Danube river fluctuations most probably come about as results of combined influences of storage effects, highly intermittent spatial behaviour of rainfall, and non-linear interaction between rainfall and the river flow \citep{Gupta1995, Kantelhardt2006, Bunde2013a}. In this paper we want to assess which of the linear or non-linear scaling properties of Danube river dynamics change and which, if any, remain invariant under a particular (dam construction) anthropogenic influence.
	
	This paper is organized as follows: Section 2 presents with a brief overview of sources of our data, and of the general (without explanations of details of procedures) methodological framework of DFA, WTS, WTMM and 20CR (re)analyses. In sub-sections 3.1 and 3.2 we present results of our usage of DFA2 and WTS to study changes in scaling of the Danube and Danube tributaries level and flow datasets that are induced by damming. In sub-section 3.3 we present results of the DFA2-WTS analysis of the 20CR reanalysis precipitation series for the dams area, and overview possible links of natural climate cycles to the enhancement or introduction of cycles that we observed in sub-sections 3.1 and 3.2. We end our paper with a list of conclusions and suggestions for future work in Section 4.
	
	\section{Data and Methods}
	
\begin{figure}[h]
		\centering\includegraphics[scale=0.7]{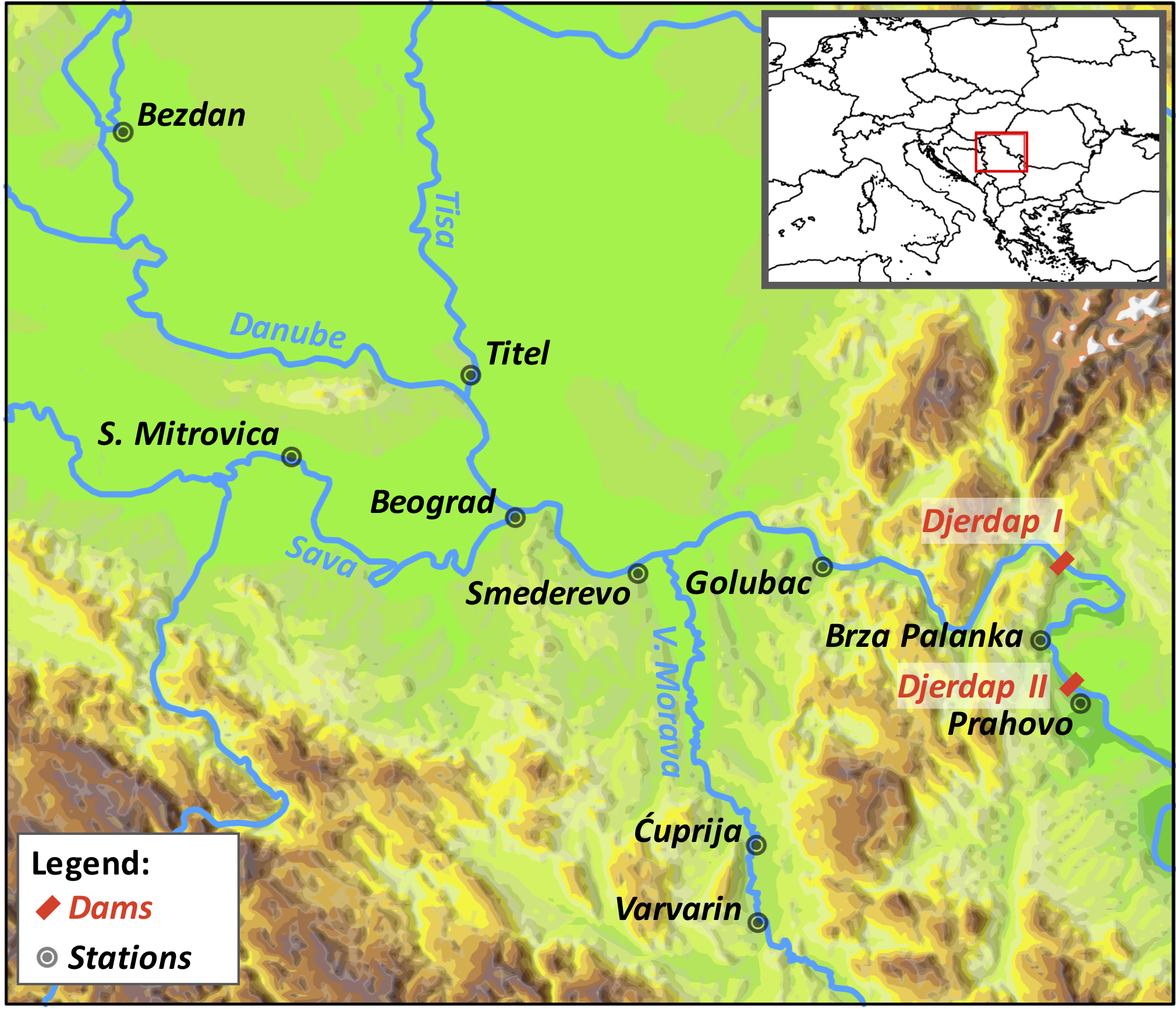}
		\caption {Hydrological stations, used in this study, and dam positions. Presented region position in wider geographical area is given as a red rectangle, in the upper right box.} 
		\label{Fig1}
	\end{figure}

	Records of daily Danube and Danube tributaries activity were provided by the Serbian Hydrometeorological Service (RHMZS,  \citet{RepublicHydrometeorologicalServiceofSerbia}). We were able to find hydrological stations with long enough historical records in the area of Djerdap/Iron Gates dams, out of which we selected to analyze data from hydrological stations (see Figure \ref{Fig1}) near the town of Golubac, situated upstream from both dams (99 km upstream from DAM1, and right bellow the accumulation lake, denoted UP in the rest of the text), the town of Brza Palanka, situated in between the two dams (59 km downstream from DAM1, denoted MD from here in), and the town of Prahovo, situated downstream from both dams (2 km downstream from DAM2, denoted DS in this paper). For these three stations only historical long records of river level were however available, thus we performed scaling analysis on the deseasoned river level data that we derived from these records \citep{Janosi1999, NunesAmaral1997}. It should be noted that even if the river level and the river flow are closely dependent variables, connected by the relationship represented in a rating curve describing the cross-section of the river \citep{Herschy, Fenton2001}, the uniqueness and the accuracy of this relationship is not exactly straightforward \citep{Fenton2001}, and in this paper we considered these two quantities as different \citep{Dahlstedt2005}. We used the notation $h$ for the river level and $q$ for the river flow. All the data we used are daily records or daily averages. 
	
	Together with data from these three main (for this paper) stations, we analyzed records from two other hydrological stations in the Danube Serbian basin, and from stations belonging to the basin of Danube tributaries Velika Morava, Sava and Tisza, with mouths in the relative vicinity of the dams. Details of the stations positions and of the elements of measurements and observations for each station data used in this paper are given in Table \ref{tab:table1}.
	
	\begin{table}[h]
		\begin{center}
			\caption{Recording stations positions, with the elements of measurements and observations for each station data used in this paper.}
			\label{tab:table1}
			\begin{tabular}{l|c|c|c|r} 
				station (basin) & km from  & km$^2$ basin  & start recording & notation\\
				& DAM1/mouth & area & year: level, flow\\
				\cline{1-5}
				\multicolumn{5}{l} {Djerdap/Iron Gates dams Danube area} \\
				\hline
				Golubac (Black Sea) & 99/1042 & 571951 & 1925, -- & UP\\
				Brza Palanka (Black Sea) & -59.2/883.8 & 576527 & 1933, -- & MD\\
				Prahovo (Black Sea) & -82/861 & 577085 & 1935, -- & DS\\	
				\cline{1-5}
				\multicolumn{5}{l} {Other Danube stations} \\
				\hline	
				Smederevo (Black Sea) & 173.23/1116.23 & 525820 & 1921, 1946 & D1\\
				Bezdan (Black Sea) & 482.59/1425.59 & 210250 & 1920, 1924 & D2\\
				\cline{1-5}
				\multicolumn{5}{l} {Tributary Velika Morava stations (mouth approx. 170km from DAM1)} \\
				\hline
				\'Cuprija (Danube) & 145.41 & 32561 & 1923, 1948 & VM1\\
				Varvarin (Danube) & 177.22 & 31548 & 1924, 1924 & VM2\\
				\cline{1-5}
				\multicolumn{5}{l} {Tributary Sava stations (mouth approx. 220km from DAM1)} \\
				\hline	
				Belgrade (Danube) & 0.82 & 95719 & 1920, -- & S1\\
				Sremska Mitrovica (Danube) & 139.24 & 87996 & 1948, 1926 & S2\\
				\cline{1-5}
				\multicolumn{5}{l} {Tributary Tisza stations (mouth approx. 270km from DAM1)} \\
				\hline
				Titel (Danube) & 8.7 & 157174 & 1930, 1965 & T1\\
				\hline
			\end{tabular}
		\end{center}
	\end{table}

	We analyzed deseasoned daily river level and (where available) river flow data. We eliminated a strong influence of a seasonal trend in both mean and variance of records, by calculating departures $r_i=\sqrt{(R_i-\bar{R})/(\bar{R_i^2}-\bar{R_i}^2)}$, where $\bar{R_i}$ is the mean value for the particular date $i$ over all years in the record \citep{Kantelhardt2003a, Livina2011, Bunde2013a}. Leap days were included in our deseasoned record $r_i$. We performed scaling analysis on the $r_i$ datasets and compared results for the three distinct time periods: a) the time period from the beginning of recording for the particular station and the year 1969, before the construction of dams (pre-development), b) the time period from 1973 to 1983, after initiation of operations of the first dam and before construction of the second dam (post-development for DAM1), and c) the time period from 1985, after both dams were operational (post-development for DAM2).
	
	In this paper, we firstly described scaling properties of river level and river flow records by calculating the scaling (or Hurst) exponent $\alpha$. To determine $\alpha$, we used the 2nd order detrended fluctuation analysis (DFA2), which (among other) systematically removes linear trends in data; detrended fluctuation analysis (DFA) was introduced as an appropriate scaling analysis to deal with nonstationary records that contain some trends of unknown form \citep{Peng1994}. In DFA, the procedure of detrending was devised so as to eliminate such trends. Resulting remarkable performance of this method in data analysis critically stems from this highly effective detrending solution, as shown by numerous systematic studies \citep{Hu2001, Chen2002, Chen2005, Xu2005, Bashan2008}. Recently, a new mathematical insight was provided that further explores how DFA operates on non-stationary data series with non-stationarity due to their (unknown) intrinsic dynamics \citep{Holl2016}. We will not explain in detail the DFA2 procedure here - for theoretical and procedural specifications we would refer to original articles that introduced the DFA procedure \citep{Peng1994} and different orders of DFA (DFAn) \citep{Kantelhardt2001}, or some of our previous utilizations of DFA2 \citep{Blesic1999, Milosevic2002}. 
	
	The advantages of using DFA over the more conventional statistical approaches (such as the calculation of autocorrelation functions or of the Fourier power spectra), for the analysis of records from any complex natural system, stem from the method design: DFA takes any typical time-dependent discrete natural data series, likely non-stationary and with unknown trends, and, by way of subtracting local trends at different time window lengths, produces a series that fluctuates much less than the original, and still has the same statistical properties \citep{Stanley2000a}. Moreover, direct calculations of conventional statistical functions (autocorrelation function or the Fourier power spectra) are hindered by the level of noise present in a typical natural record, caused by possible non-stationarities in the data. DFA calculates the fluctuation function, by definition a sum over autocorrelations \citep{Holl2015b}, and thus fluctuates less; one therefore uses a function that is entirely defined by the autocorrelation function, but is markedly more stable \citep{Bunde2013}. In the case of long-range autocorrelated data the DFA function $F(n)$, due to the inherent power-law data dynamics, presents as a straight line on log-log graphs of dependance of $F(n)$ of the time scale $n$, allowing for quantification of scaling by the corresponding power-law exponent (log-log slope) $\alpha$. 
	
	If the time series analyzed is short-range autocorrelated, or has no correlations at all, $F(n)$ behaves as $n^{1/2}$ \citep{Peng1994}. Differently, for data with power-law long-range autocorrelations $F(n)\sim n^{\alpha}$, with $\alpha \neq 0.5$. Records are generally labelled long-range autocorrelated, or long-term persistent (LTP), when the corresponding autocorrelation function $C(s)$ decays by a power law $C(s)\sim s^{-\gamma}$, for $s > 0$ and $N\rightarrow \infty$, and when the mean correlation time, defined as $T=\int_{0}^{\infty} C(t)dt$ diverges \citep{Holl2015b}. The exponent $\gamma$ can then be used to quantify the nature and the level of autocorrelations in the record; for stationary cases $\gamma$ lies in the range $0<\gamma<1$. It has been shown that, in this case, the Fourier power spectral density decreases as a power law as well, with $E_F(\omega)\sim \omega^{-\beta}$, and the exponent $\beta$ in the range $-1<\beta<1$ \citep{Peng1993a}. The exponent $\alpha$, associated with the detrended fluctuation function $F(n)$, can be related to both $\gamma$ and $\beta$ through scaling relations $\alpha=1-\gamma/2$ and $\alpha=(\beta+1)/2$ \citep{Peng1993a}. This bounds $\alpha$ to a range $0<\alpha<1$ for stationary records, where $0.5 <\alpha<1$ indicates that the record is long-term persistent. Instances when $\alpha \geq 1$, that will be of interest to the dataset used in this paper, imply the existence of intrinsic non-stationarities in autocorrelated data \citep{Holl2016}. When this is the prevalent data dynamics, the corresponding DFA functions exhibit crossovers, while $\alpha \geq 1$ may mean that the underlying process is of a composite nature \citep{Holl2015b}, or that there exists an imbalance between different noise inputs \citep{Hausdorff1996}. Finally, $\alpha = 1.5$ quantifies Brown noise, the integration of white noise.
	
	A pure LTP behaviour rarely occurs in natural records, and thus the corresponding DFA2 functions, depicted on log-log graphs, are rarely ideal linear functions. Instead, they tend to contain transient crossovers in scaling that stem from occurrences of irregular phenomena of different types \citep{Mallat1992, Hu2001}, most commonly from the effects of mixtures of cyclic components that locally perturb DFA2 functions \citep{Mandelbrot1969}. It has been shown \citep{Hu2001} that the spread of any such perturbation and the length of scales that it covers until asymptotically resuming to the DFA behaviour dominated by LTP noise depends on the scaling exponent $\alpha$ and the period and/or amplitude of the hypothetical periodic trend, and is generally much less visible for greater values of $\alpha$ (see detailed theoretical explanations in \citet{Mandelbrot1969} and \citet{Hu2001}). When effects of such irregularities are visible on DFA2 curves, but are not comparatively strong to change global behaviour of DFA2 functions, we used wavelet transform spectral analysis (WTS) to investigate them in detail. We use notation DFA2-WTS for the results independently derived from DFA2 and WTS in this paper. 
	
	The wavelet transform (WT) method was introduced in order to circumvent the uncertainty principle problem in classical signal analysis \citep{Stratimirovic2018} and achieve better signal localization in both time and frequency than that of the classical Fourier transform approaches \citep{Morlet1982, Grossmann1984}. In WT, the size of an examination window is adjusted to the frequency analyzed; in this way an adequate time resolution for high frequencies and a good frequency resolution for low frequencies is achieved in a single transform \citep{Bracic1998}. For detailed explanation of the WTS procedure used in this paper please see e.g. \citet{Morlet1982}, \citet{Grossmann1984}, \citet{Astafeva1996}, or previous uses by our group (including the introduction to the multifractal data analysis) in \citet{Stratimirovic2001}. In order to obtain the kind of results comparable with those of the DFA2 method, in this paper we calculated the mean wavelet power spectra $E_W(n)$, which can be related \citep{Perrier1995a} to the corresponding Fourier power spectra $E_F(\omega)$ via the formula that assigns that if any of the two spectra - $E_W(n)$ or $E_F(\omega)$ - exhibit power-law behavior, then the other will be of the power-law type as well, with the same power-law exponent $\beta$ \citep{Stratimirovic2001}. This practically means that the function of the wavelet scalegram is the same as that of the classical Fourier spectrum: it calculates the contribution to the signal energy along the scale of $n$. We standardly use Morlet wavelets of the 6th order as a wavelet basis for our analysis. Morlet wavelets provide with an optimal joint time-frequency localization \citep{Goupillaud1984, Torrence1998} and are particularly well adapted to estimate local regularity of functions \citep{Mallat1992}. In local power spectra the Morlet wavelet is narrow in spectral (scale)-space, and broad in the time-space, which produces very well localized, relatively sharp peaks in global WT spectra, the averages of local spectra over time \citep{Torrence1998}. This allows for the reliable determination of locations and spatial distributions of both periodic or non-periodic cycles and significant singular events in non-stationary time series \citep{Mallat1992}. To assess the significance of obtained cycles - peaks in WT spectra, we utilized statistical significance testing proposed in \citet{Torrence1998}, against the analyzed signal as the noise background, as explained in \citet{Stratimirovic2018}. 
	
	Following what has been proposed earlier \citep{Pandey1998, Kantelhardt2003a, Koscielny-Bunde2006, Kantelhardt2006, Bunde2013a}, and the fact that river records are detectable manifestations of complex underlying hydroclimatic processes which could also give rise to multifractal behaviour \citep{Stanley2002}, in this paper we apply the multifractal formalism to study multiscaling properties of Danube data. We used formalism adapted for time-series analyses performed by wavelet transform method - the so-called wavelet transform modulus maxima (WTMM) method; for details see, e.g., introduction in \citet{Arneodo1995} and detailed description in \citet{Stratimirovic2001}. We used WTMM to obtain distributions of the singularity spectra $D(H)$, related to the fractal dimensions of the analyzed time series \citep{Stratimirovic2001}. For multifractal series $D(H)$ is a parabolic curve whose maximum position on the x-axis indicates the value of the (monofractal, or global) H\"{o}lder exponent $H$ of the series, with $H=\alpha-1$ \citep{Scafetta2003}. For monofractal data, or time series that can be described by a single H\"{o}lder exponent, $D(H)$ collapses to a single point.
	
	Finally, in order to assess the association of changes in long-term properties of river dynamics with the local climate and climatic characteristics, we performed analysis of daily precipitation averaged over catchment area upstream from dams. Precipitation is obtained from NOAA-CIRES 20th Century Global Reanalysis Version 2 \citep{Compo2011}, with horizontal resolution of $2^{\circ}$ in both longitudinal and latitudinal direction. For the area of interest 20 points from global reanalysis domain were extracted and averaged in space, for the time period from 1871 to 2012.
	
	\section{Results}
	
	\subsection{River Level Dynamics in the Vicinity of Dams}
	
\begin{figure}[h]
		\centering\includegraphics[scale=0.5]{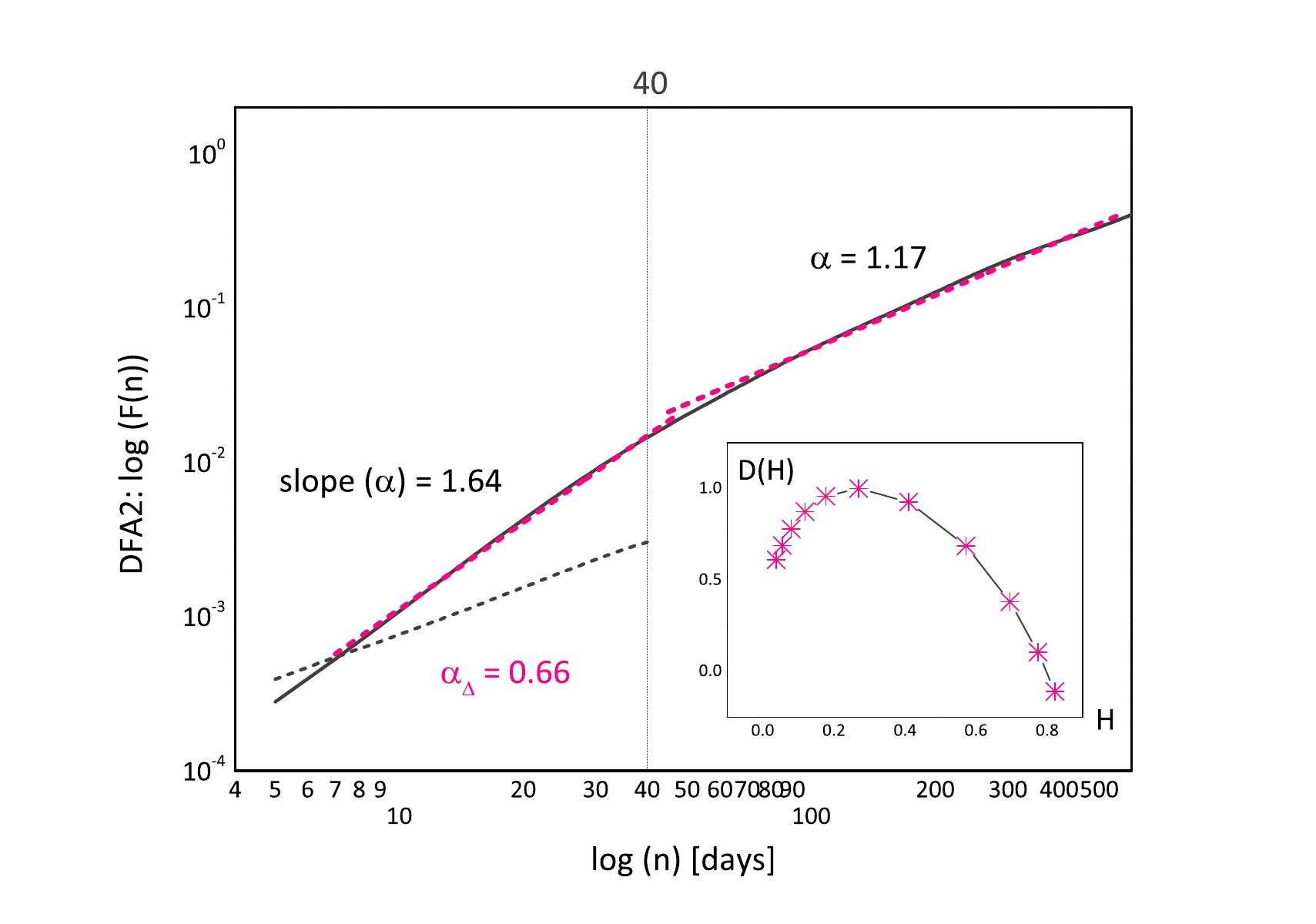}
		\caption {Typical result of the DFA2 analysis of a time series of Danube river level data from hydrological station in the vicinity of Djerdap/Iron Gates dams. DFA2 function (solid line) is given on a log-log graph, together with linear fits below and above the crossover in scaling (pink dashed lines), and the DFA2 function of the corresponding series of increments, in the small scales region (dashed line). Values of exponents $\alpha_1$ and $\alpha_2$ are provided; for the estimation of errors to DFA2 exponents see \citet{Bashan2008}. Inset: Singularity spectrum $D(H)$ of the WTMM method, calculated for the time series depicted in this Figure.} 
		\label{Fig2}		
	\end{figure}

	In Figure \ref{Fig2} we present a typical result from our DFA2 analysis. In all the level and flow records from hydrological stations on the four rivers (that is, Danube and its three tributaries) that we investigated we found that DFA2 curves are approximately straight lines in log-log plots; this is in accordance with previous research. We found crossover in DFA2 behaviour of our records, which indicate that analyzed data dynamics is nonstationary and probably stems from a composition of dominant influences \citep{Hausdorff1996, Holl2016}. We found that the scaling that we observed exhibits crossover at timescales of several weeks (in the range of 15 to 40 days), with scaling exponents $\alpha_1$ slightly above 1.5 at time scales below the crossover region, indicating very strong short-term correlations in the small scales area, in accordance with previous studies. For this region of scales we performed a test of autocovariance difference, prescribed by \citet{Holl2016} to assess whether values of $\alpha_1 > 1$ in this time region are due to the existence of intrinsic non-stationarities in the data. Our data did not show the autocovariance difference and are thus meeting this criterion. It has been proposed by numerous previous studies that, in such a case, these types of strong autocorrelations should be assessed through investigations of fluctuations in series of increments of the original series studied, that is in the dataset that consists of values of $\triangle x_i=x_{i+1}-x_i$ \citep{Koscielny-Bunde2006}. If the original record has scaling exponent $\alpha_1 > 1$, or especially if $\alpha_1 \approx 1.5$ as in the case of our river level data, the exponent of the series of increments $\triangle x_i$ would be $\alpha_{\triangle} = 1-\alpha_1$; an example of the DFA2 result for the corresponding series of increments, in the small scales region, is given in Figure \ref{Fig2}. Finally, we found that above the crossover all our data exhibit autocorrelated, or persistent behaviour, with the scaling exponents $\alpha_2$ smaller than in the region below the crossover, but still in most cases with $\alpha_2 > 1$, which indicates that our records stay in nonstationary regime also in the longer scales time range (see Figure \ref{Fig2}). This result is somewhat in contrast with some of the observations of scaling in river flow data, including data from river Danube \citep{Koscielny-Bunde2006, Kantelhardt2006}, and may be manifestation of the observation that for the same river, the scaling exponent $\alpha$ may increase down the river, when the basin size increases \citep{Bunde2013a}. Finally, we investigated the possible multifractal origin of our non-stationary records: in all the cases analyzed we found, as depicted in the inset of Figure \ref{Fig2}, that the WTMM singularity spectra $D(H)$ manifest in rather broad parabolic curves, signs of their multifractality \citep{Koscielny-Bunde2006, Kantelhardt2006}.  
	
\begin{figure}[h]
		\centering\includegraphics[scale=0.5]{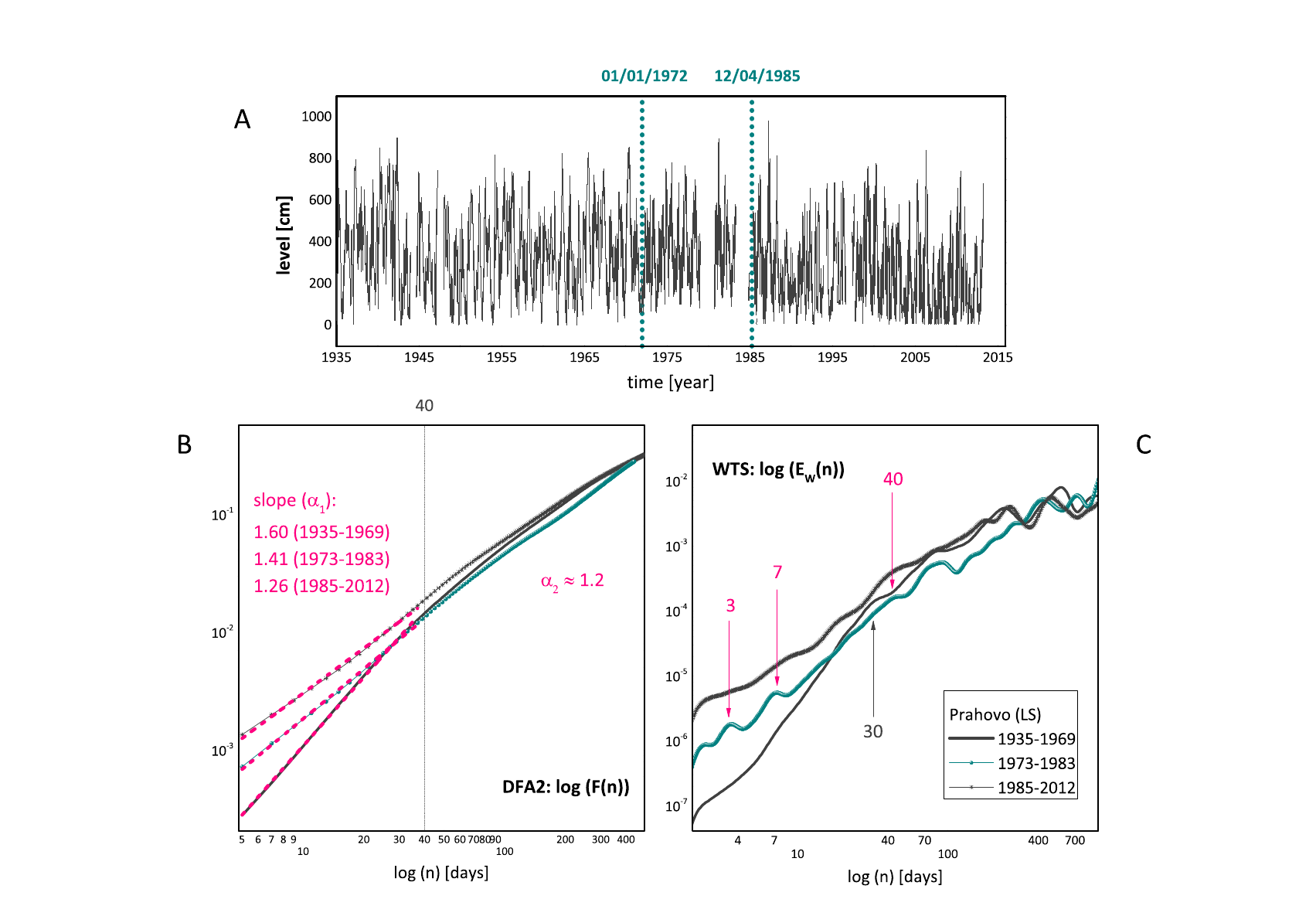}
		\caption {Results of the DFA2-WTS analysis of the time series of Danube river level records from the hydrological station Prahovo, positioned downstream from Djerdap/Iron Gates dams. (A) Data. Vertical cyan dotted lines indicate commencement of operation for the two dams; (B) DFA2 functions for the period before construction of dams (gray solid lines), after the construction of DAM1 (cyan filled circles), and after the construction of DAM2 (gray asterisks), together with linear fits to the DFA2 curves in the region below the crossover (pink dashed lines). Values of DFA2 exponents $\alpha_1$ and $\alpha_2$ are provided, while the approximate position of the crossover is indicated by the vertical dotted line; (C) The corresponding WTS of the three time period given in (B), with significant WTS peaks that appear in the small-scales region marked with arrows. DFA2-WTS functions are given on log-log plots.} 
		\label{Fig3}		
	\end{figure}

	Our DFA2-WTS results for the hydrological stations in the vicinity of dams show markedly different scaling that depends on the position of the station in relation to dams, and on the time period investigated. This is how, for the hydrological station downstream from both dams (labelled DS), we found a change of scaling dynamics in the periods after construction of dams that is evident only in the region below the crossover point. There, even if no significant change in the dynamics of the river level with the construction of dams is visible from the raw data (see Figure \ref{Fig3}A), the value of the small-scales DFA2 scaling exponent $\alpha_1$ is significantly lowered in the period after the construction of DAM1, and this effect visibly reappears after the construction of DAM2 (see Figure \ref{Fig3}B). The WTS power spectra, depicted in Figure \ref{Fig3}C, show that these changes come about for the appearance of new cycles in the short-scales region with the amplitudes that notably influence scaling below the crossover, that is, decrease the scaling exponent there. These cycles appear at periods at approximately 2 days, at 3 days, at 7 days and at approximately 40 days; the 40-days cycle may be a result of prolongation of the natural 30-day cycle, visible in the pre-construction period data, and it also coincides with the (new) position of the crossover in scaling. After the crossover the DFA2 functions for all three investigated time periods have the same value of slope of $\alpha_2 \approx 1.2$.
	
\begin{figure}[h]
		\centering\includegraphics[scale=0.5]{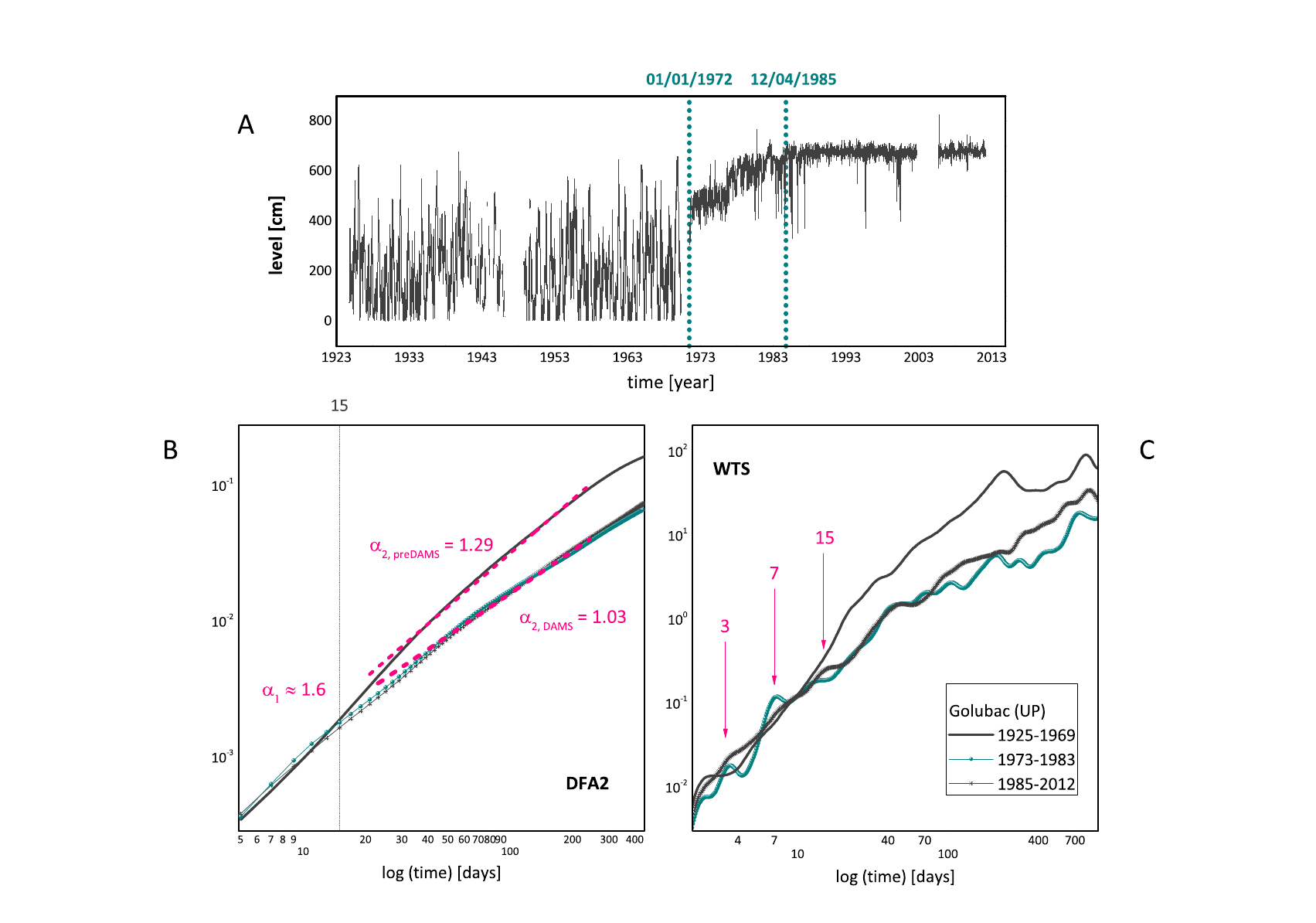}
		\caption {Results of the DFA2-WTS analysis of the time series of Danube river level records from the hydrological station Golubac, positioned upstream from Djerdap/Iron Gates dams. (A) Data. Vertical cyan dotted lines indicate commencement of operation for the two dams; (B) DFA2 functions for the period before construction of dams (gray solid lines), after the construction of DAM1 (cyan filled circles), and after the construction of DAM2 (gray asterisks), together with linear fits to the DFA2 curves in the region above the crossover (pink dashed lines). Values of DFA2 exponents $\alpha_1$ and $\alpha_2$ are provided, while the approximate position of the crossover is indicated by the vertical dotted line; (C) The corresponding WTS of the three time period given in (B), with significant WTS peaks that appear in the small-scales region marked with arrows. DFA2-WTS functions are given on log-log plots.} 
		\label{Fig4}		
	\end{figure}

	Figure \ref{Fig4} displays results we obtained for the records from hydrological station upstream from both dams (denoted UP), showing how the change in dynamics upstream from dams is quite opposite to the one presented in Figure \ref{Fig3}. Namely, here the change of river level dynamics is already visible from the raw data (given in Figure \ref{Fig4}A), with the visible increase in water level after the construction of DAM1 and additional increase after the construction of DAM2, coupled with substantial reduction of water level variability. This is manifested in visible changes in DFA2 scaling (given in Figure \ref{Fig4}B) in the large scales area, above the crossover, while DFA2 scaling remains unaltered in the small-scales region. The corresponding WTS curves, given in Figure \ref{Fig4}C show that, even if new cycles at approximately 2 days, at 3 days, at 7 days and at 15 days do appear in the small-scales region of WTS spectra after the construction of dams, they do not seam to alter scaling dynamics there. The scaling after the crossover, now positioned at approximately 15 days, is thus rather being altered by lowering or even loss of cyclic influence and noise at larger scales.
	
\begin{figure}[h]
		\centering\includegraphics[scale=0.5]{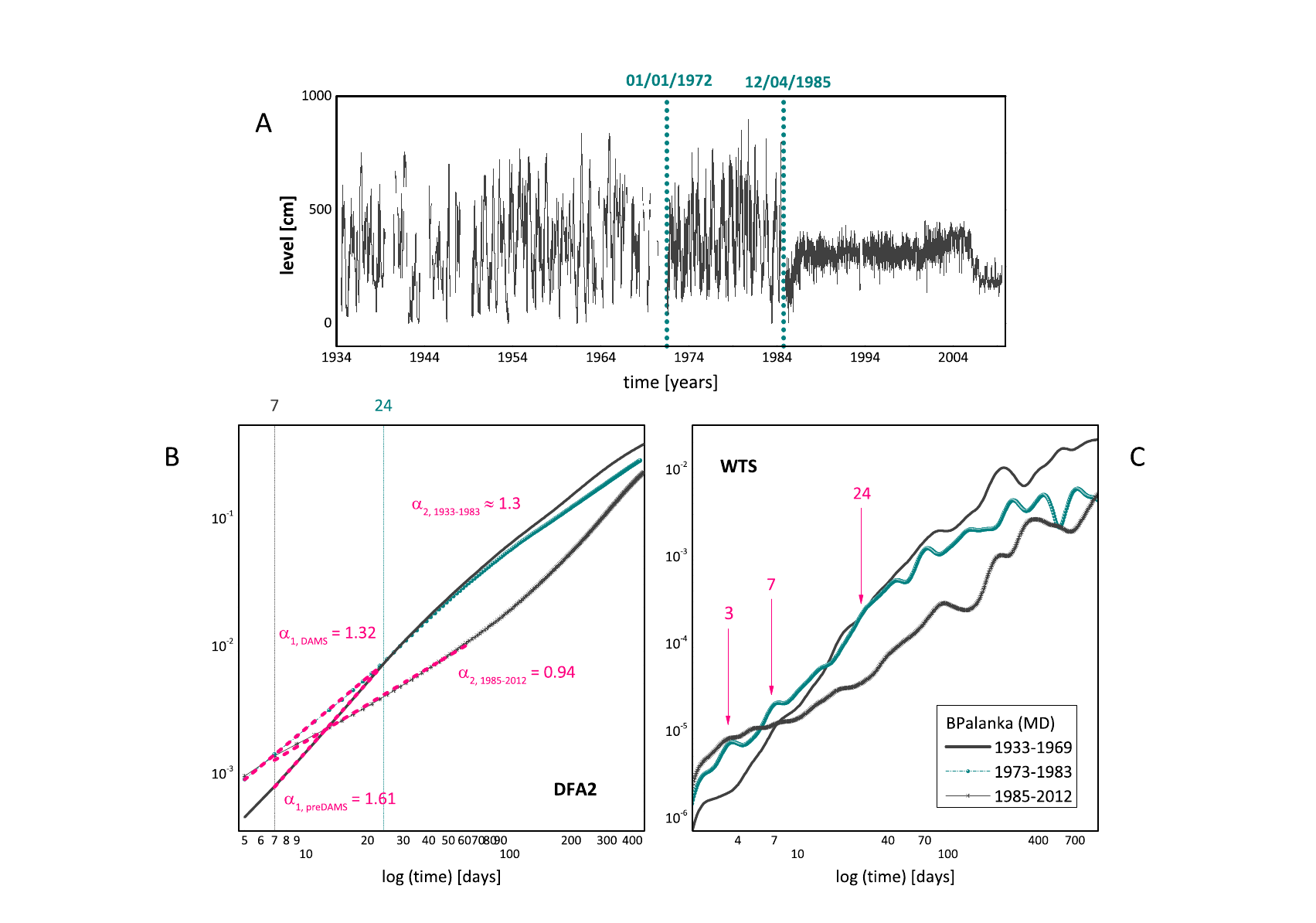}
		\caption {Results of the DFA2-WTS analysis of the time series of Danube river level records from the hydrological station Brza Palanka, positioned in between the two Djerdap/Iron Gates dams, depicted in the same way as results in Figures \ref{Fig3} and \ref{Fig4}.} 
		\label{Fig5}		
	\end{figure}

	Finally, in Figure \ref{Fig5} we show the extraordinary mixture of the two behavioural changes in scaling depicted above, for the records from the hydrological station Brza Palanka that is situated between two dams. After the construction of DAM1 this station is positioned downstream from the dam and its DFA2-WTS functions change like in the graphs presented in Figure \ref{Fig3}: only the scaling exponent $\alpha_1$ changes, decreasing in value due to the effects of new and/or enhanced cycles at approximately 2 days, at 3 days, at 7 days and at approximately 24 days (the last cycle also delineates the position of the crossover). After DAM2 is constructed the scaling in this station, for it is now situated upstream from the new dam, changes in a way similar to the one depicted in Figure \ref{Fig4}: DFA2 function has a crossover at a very short scale, after which its scaling exponent $\alpha_2$ decreases significantly due to the drop in amplitude of larger-scale cycles and noise. This behaviour finally changes again at the very large scales, where low-frequency noise seams to dominate the scaling.
	
\begin{figure}[h]
		\centering\includegraphics[scale=0.5]{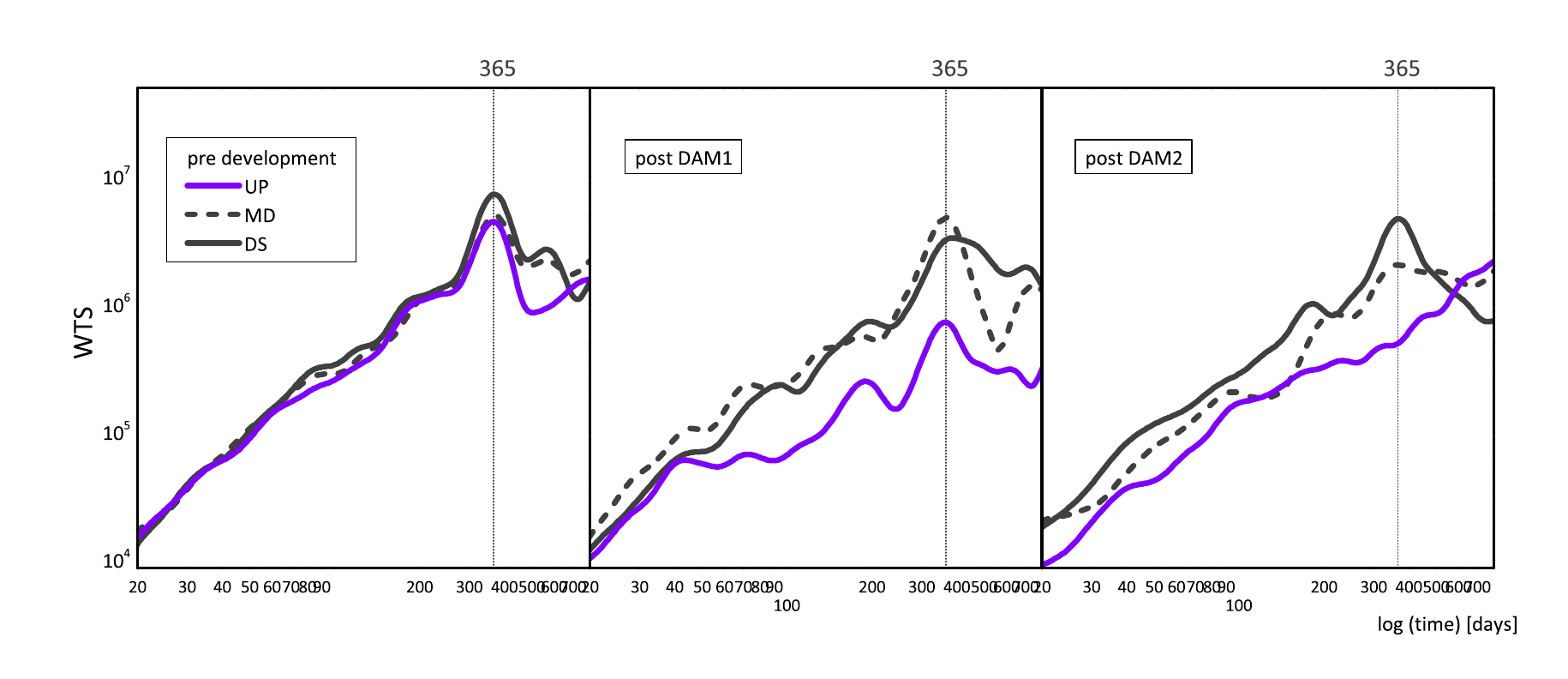}
		\caption {WTS graphs of the original (not deseasoned) records of the three hydrological stations for the three time periods related to the times of construction of Djerdap/Iron Gates dams. Dotted vertical lines at $t=365$ days are given as visual guides. Partial destruction and the complete loss of the annual cycles are visible for stations MD (dashed line) and UP (violet line) in the time periods after the construction of dams. } 
		\label{Fig6}		
	\end{figure}

	In order to be able to study more closely the effect of decrease of influence of large-scale natural cycles and noise in upstream water level after construction of dams, and particularly in order to be able to examine modifications in the annual cycle of water level caused by damming, we calculated WTS power spectra for the original records of the three stations, as they are before deseasoning. It is important to note here that we used these calculations only to look into the behaviour of annual cycles, and not to (re)consider scaling properties. Namely, it has been shown repeatedly by other groups and by us that, when the original records are used, in the range of scales of our interest, the seasonal trend dominates DFA2-WTS behaviour in such a profound way that the accurate estimation of scaling is impossible (see, for example, \citet{Mandelbrot1969}, \citet{Hu2001}, or \citet{Blesic2018}). In Figure \ref{Fig6} we present results of the WTS analysis of the original (not deseasoned) data. We present WTS functions for the three hydrological stations in time periods before construction of dams, after the construction of the first dam, and after both dams were operational. Figure \ref{Fig6} clearly shows that the construction of dams is connected to the loss of amplitude of large frequency cycles and noise for records taken upstream from dams, which was already apparent from results given in Figures \ref{Fig4} and \ref{Fig5}, and that this is particularly connected to not just a decrease in amplitude, but even loss of the annual cycle in analyzed data. In order to check whether this loss is maybe prominent but in a way temporary, we did a separate analysis of the data for the UP and MD stations for the period of 2002 to 2012 (last ten years we had in record), and we found the same WTS behaviour, that is, we found that the annual cycle is still completely (in the UP station) or partially (in the MD station) destroyed (these results are placed in the online supporting material to this paper).
	
	\subsection{Influence of Damming on the Level and Flow Dynamics of Upstream Danube and Danube Tributaries' Stations}
	
	We analyzed data from seven hydrological stations positioned on river Danube or its three upstream tributaries. Of those the station Smederevo (D1 in our notation) is, at approximately 170 km, the closest to the dams area, while the station Bezdan (D2; 480 km away) is the most distant. For all but one of these stations we had access to the river flow data in addition to the river level records, which gave us the opportunity to compare dynamics of these two variables to some extent.
	
	Of all the upstream Danube and Danube tributaries stations in our dataset, we found changes in scaling dynamics that can be connected to damming in three sets of hydrological records: in Danube upstream station Smederevo (D1 in our notation), in Velika Morava tributary \'Cuprija (VM1) station, and in the river Sava tributary Belgrade (S1) station. This bounds the range of effects of damming to approximately 220 km upstream from the damming area (distance of S1 from dams).
	
	In these three stations we found visible change of behaviour in the raw river level data, but not in the raw river flow data. Out of these, in station \'Cuprija (VM1) we found visible decrease of the river level, while in stations Smederevo (D1) and Belgrade (S1), like in the station UP in the vicinity of dams, we found a visible rise of the river level as the effect of damming. Negative trend observed in VM1 was also recorded for many hydrological stations in southern part of Velika Morava catchment, and may partially be explain by negative precipitation trend in southern and southeastern parts of Serbia \citep{Dimkic2018}. In all of the three remote stations we found the change in the water level accompanied with the visible decrease of level's variability, as in stations UP and MD in the vicinity of dams. In all of the three historical records - D1, VM1, and S1, we were able to only see changes induced by damming in the WTS functions of the original (before deseasoning) river level data, while the DFA2-WTS results of both the deseasoned river level and river flow data remained unaffected over time. In addition, scaling (that is DFA2-WTS) dynamics of the (deseasoned) river level and river flow records was virtually the same, within the range of error, in all of the analyzed data, with the values of scaling exponent $\alpha_2$ above the crossover ranging from 0.85 to 1.1. We provide an example of these results for station \'Cuprija (VM1) in supporting material to this paper. 
	
\begin{figure}[h]
		\centering\includegraphics[scale=0.5]{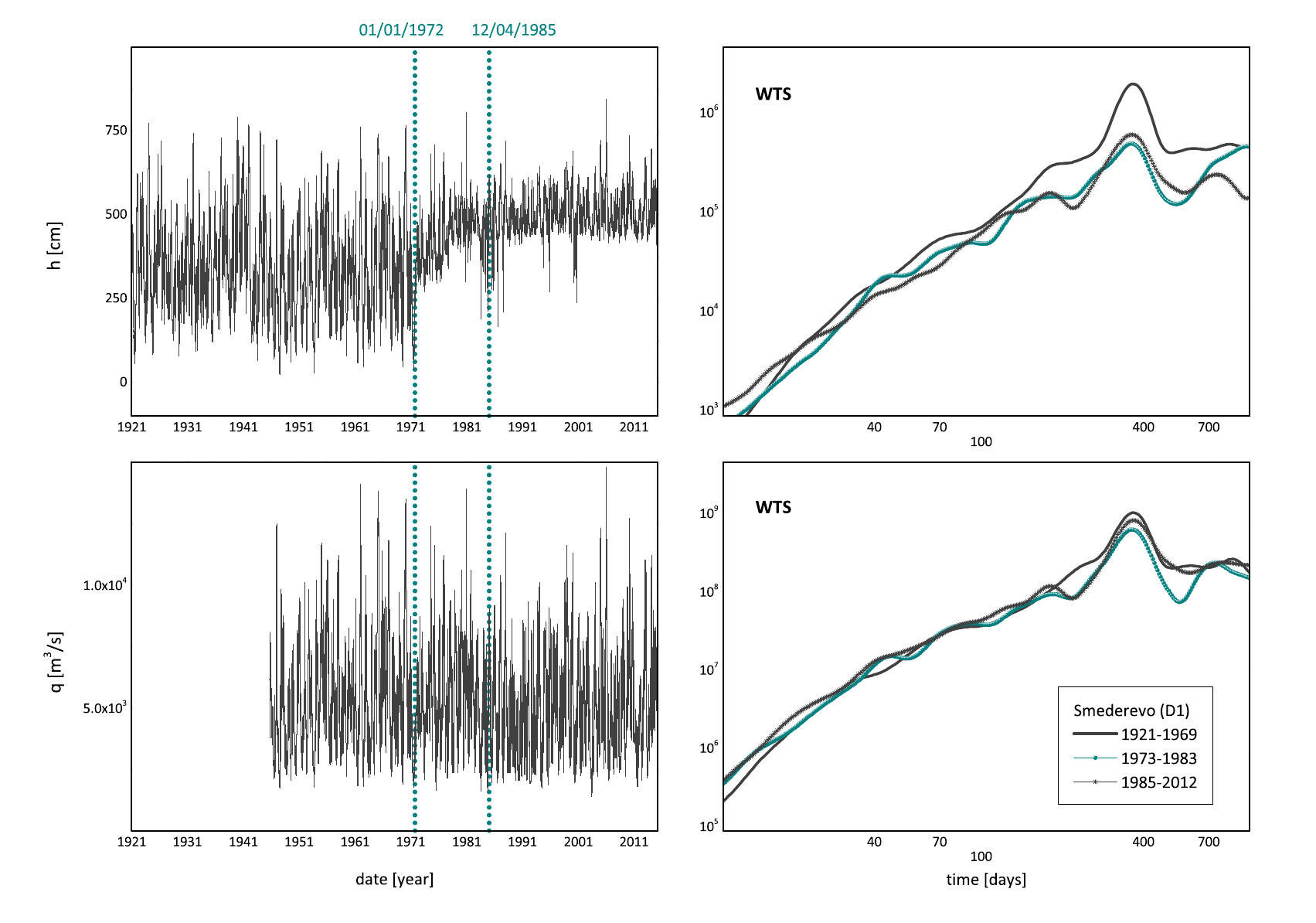}
		\caption {River level historical records (upper raw, first column), starting in 1921, and their WTS functions (upper raw, second column) for the three time periods, together with the river flow historical records (lower raw, first column), starting in 1946, and their WTS functions (lower raw, second column), for the three time periods, for the hydrological station Smederevo (D1) positioned on river Danube, 173 km upstream from the dams. Change in the amplitude of WTS annual cycles, as effect of damming on the river level, is visible in the upper WTS graph. } 
		\label{Fig7}		
	\end{figure}

	In Figure \ref{Fig7} we provide graphs of the raw river level and the river flow records, together with their (before deseasoning) WTS functions, for hydrological station Smederevo (D1). In this record, as in records from the hydrological stations VM1 and S1, the effects of damming are visible in the change of the river level accompanied by the decrease of its fluctuation dynamics in the raw data, and in the decrease of the amplitude of the annual cycle, in their original (before deseasoning) WTS functions.
	
	\begin{figure}[h]
		\centering\includegraphics[scale=0.5]{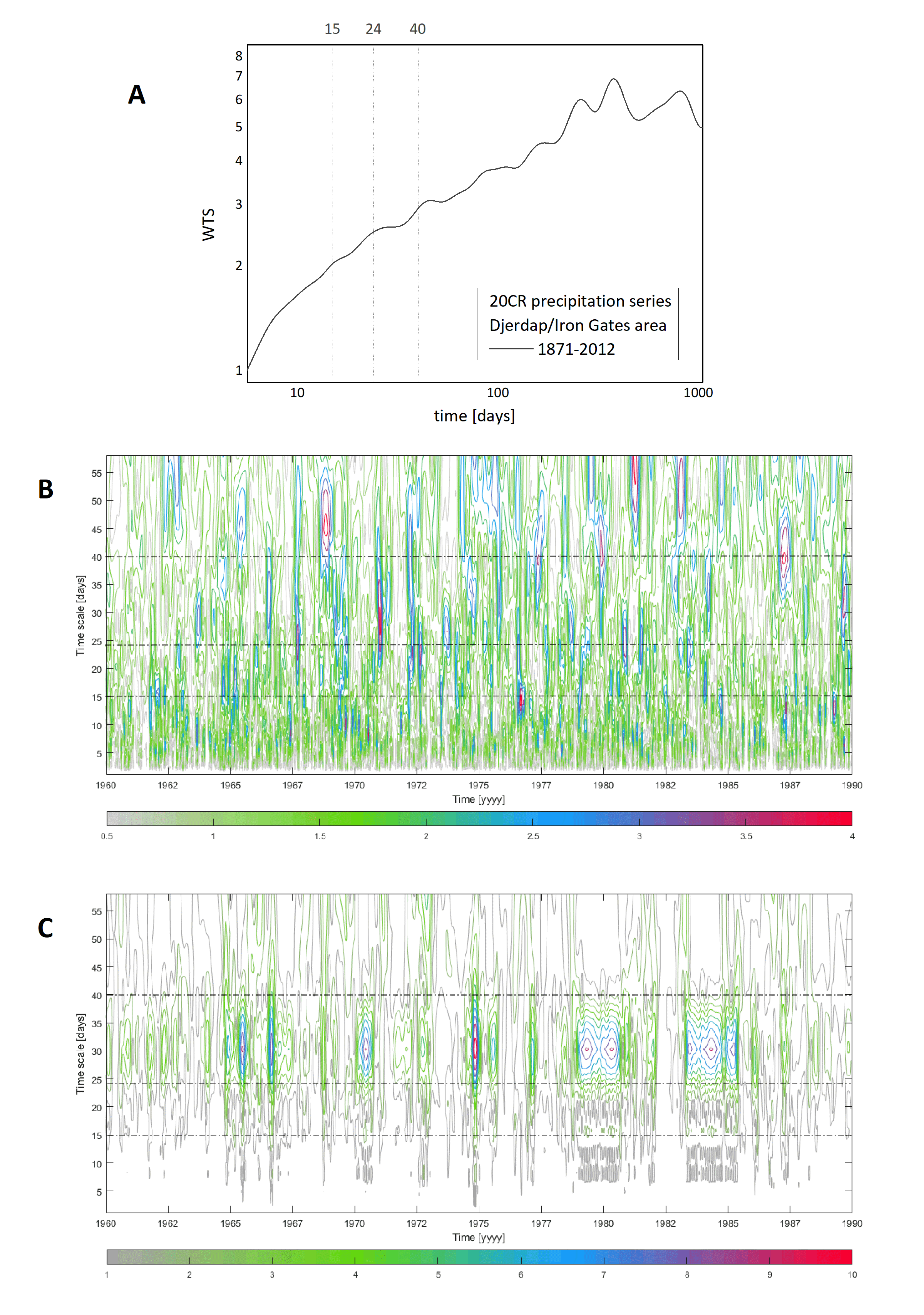}
		\caption {(A) WTS of a 20CR time series of precipitation in the period 1871-2012 in the Djerdap/Iron Gates geographical area. Vertical dotted lines at 15 days, 24 days and 40 days are given as visual guides. (B) The pattern of WT coefficients for the same 20CR time series depicted in (A), for the period 1961-1990. Horizontal dotted lines at 15 days, 24 days, and 40 days are given as visual guides. The colorbar codes the increase of the intensity of absolute values of WT coefficients. (C) Same display as in (B), for the time series of Prahovo (DS) station river level records, for the period 1961-1990.  } 
		\label{Fig8}		
	\end{figure}
	
	Finally, the WTMM multifractal analysis we performed on all our records showed, as depicted in the inset of Figure \ref{Fig1}, distributions of $D(H)$ as broad parabolic curves, signs of underlying multifractality and rich structure of time series \citep{Stosic2015}, with slightly left-skewed shapes that speak of dominance of fractal exponents that describe the scaling of large fluctuations \citep{Stosic2015}, in the periods before construction of dams. After damming we found that the shapes of $D(H)$ distributions changed for UP and MD stations, where we found visible decrease of the $D(H)$ width and change from asymmetric into a more symmetric shape. This would mean that the river level dynamics in the upstream stations in the vicinity of dams has lost the richness of its multifractal structure after damming, probably due to the loss of influence of some of the larger-scale phenomena. We did not find changes consistent with these in our other stations, after the damming. We present our WTMM results in the supporting material to this paper.
	
	\subsection{Changes in River Level Data Associated with Climate}
	
	Since the major component of the hydrological cycle that determines river level and river flow is accumulated precipitation over catchment area, we investigated precipitation regime over catchment area upstream from dams, trying to find similar cycles as the ones presented in river level data. For this purpose, we calculated daily accumulated precipitation, averaged over catchment area, from the NOAA-CIRES 20th Century Global Reanalysis (20CR) for the period 1871--2012. For this long-term time series of daily precipitation that overlaps with river level data we calculated WTS curves for deseasoned simulation series. According to analysis presented in Figure \ref{Fig8}A, the cycles of 15, 24 and 40 days that are present in river level data analysis can also be found in precipitation data. These cycles belong to intra-seasonal time scales and can be part of low-frequency variability modes in atmosphere. However, these cycles did not present as significant \citep{Stratimirovic2018} in the reanalysis data, as they did in the river level records in the vicinity of dams. Diminishing significance of cycles can be a result of the process of data averaging within the simulation cell \citep{Blesic2018}. We have, however, additionally looked into the local spectra of wavelet transforms (that is, into the temporal patterns of WT coefficients) of reanalysis data and river level records in the vicinity of dams, in order to check whether they share the timing and the sources of cycles that appear in their WTS spectra. In Figure \ref{Fig8}B and \ref{Fig8}C we show patterns of WT coefficients for the 20CR precipitation series and the Prahovo (DS) station river level record, for the period 1960--1990. What these graphs show is that, while cycles at 15, 24 and 40 days appear as non-periodic in 20CR series, probably following natural climatic influences, the rise in values of WT coefficients connected to these cycles in DS local spectrum is less stochastic and obviously human-related. This is how activities appearing in repetition on 15, 24 and 40 days intervals present in DS local spectrum around year 1965, when the damming works have started. The same pattern replicates at the end of 1970 and the beginning of 1971, when the DAM1 started operations, and in 1975, when building of DAM2 probably started. The periods of extensive reservoir and electricity production management are also visible in two additional periods - from 1979 to 1981, and from 1983 to 1986. These activities may be the result of rain forecasts which have been used by the operators to adjust the operation of dams, and thus connected to the climatic phenomena, but are visibly different, more ordered and longer in duration than any of the natural events that might have caused them.    
	
	\section {Discussion and Conclusions}  
	
	In this paper we used DFA and WTS, two methods of scaling analysis, to assess changes in long-term dynamics of Danube river level and flow associated with Djerdap/Iron Gates dams, in order to further understand and quantify those alterations, and produce explanations that could be of use in future environmental assessments. In accordance with previous similar studies, we found scaling, or presence of long-term persistence (LTP) in all our records. We found LTP to be a sign of a high non-stationarity of our data, with scaling exponents $\alpha, \beta > 1$, and presence of crossover in scaling on scales of 7--40 days. Below the crossover, we found very strong short-term autocorrelations in all our records, with DFA2 exponent values of $1.2 \leq \alpha_1 \leq 1.65$; it was reported before that these values of $\alpha_1$ indicate that the short-term (below the crossover) autocorrelations can be modelled as an ARMA processes with the characteristic autocorrelation time related to decay time of floods \citep{Kantelhardt2006}. Above the crossover, we obtained very pronounced, in most cases non-stationary LTP, with $0.85 \leq \alpha_2 \leq 1.2$ for Danube or Danube tributaries river level. This result is somewhat different (producing larger $\alpha_2$ values than reported before) from findings for Danube river flow scaling \citep{Koscielny-Bunde2006, Kantelhardt2006}, and may reflect the observation that for the same river the scaling exponent may increase down the river, when the basin size increases \citep{Bunde2013a}. This variation from previous results may additionally arise from the influence of different flood mechanisms on scaling, for it has also been shown that rain-induced floods and snow-induced floods may introduce different spatial scaling behaviour \citep{Kantelhardt2006}. Finally, previous researches \citep{Kantelhardt2006} suggest that the crossover timescale found in our records is similar to the period of planetary waves, which can influence decay time of floods and thus the river level's short-term dynamics.  
	
	Our results show visible and significant impacts of the building of Djerdap/Iron Gates dams on scaling of the Danube river level. We observed changes in scaling that affect DFA2 and WTS functions in a different way, depending on the position of the data collecting hydrological station relative to dams. For time series of river level measured downstream from the damming area we found changes in scaling in the short time-scales region, below the crossover. There, DFA2-WTS functions show visible decrease in scaling, when dynamics of records collected before building of dams is compared to the one after damming, while the scaling in large time-scales region, after the crossover, remains unchanged. This change in short-term dynamics is partly brought by the appearance or increase of amplitude of WTS cycles positioned at 2 days, at 3 days, at 7 days, and at 40 days, which are most probably related to the timing and magnitude of the controlled release of water. The appearance of these human-made or human-enhances cycles may reflect dams' working regime in electricity production (such as is probably the appearance of a 2-day cycle, reflecting the drop of production during weekends), or in protection of downstream area from flooding, in managing high flows at around 3-day and 7-day intervals, as well as possibly prolonging low flow durations from 30 to 40 days  \citep{Richter2002, Magilligan2005a, IdentificationoflongtermhighflowregimechangesinselectedstationsalongtheDanubeRiver}. If this is the case, the obtained raise in amplitude of these cycles, coupled with the loss of relative spectral contributions of other short-range noise and the decrease of values of $\alpha_1$, all confirm postulated key influence of flood mechanisms on short-term scaling of river level. This result should be further inspected for downstream hydrological stations outside of Serbia \citep{IdentificationoflongtermhighflowregimechangesinselectedstationsalongtheDanubeRiver}, due to the availability of records the geographical limit of our study, to corroborate our findings and explore the spatial range of the observed effect in short-term scaling.
	
	We additionally found a distinct effect of damming on river level scaling upstream from Djerdap/Iron Gates dams. Even if the promotion of 2-day, 3-day, 7-day, and 15-day flow pulses regulation is visible also in the upstream WT spectra, they do not seam to affect upstream scaling dynamics. Instead, here it is the long-term scaling that is affected by damming, with visible decrease of values of $\alpha_2$ in our DFA2 functions. Our WTS findings show that this change is mainly brought by the complete (in the vicinity of dams) or partial (further upstream) loss of the natural annual cycle, together with decrease of amplitude of other larger-scales noises. This is a dramatic alteration of upstream river level dynamics. It provides information that, in addition to obvious transformation upstream of dams from a free-flowing river ecosystem to an artificial reservoir habitat \citep{InternationalRivers2019}, a major change in the river hydrodynamics also occurs as an effect of damming. Our findings additionally inform that seasonality is a dominant source of river level's long-term scaling, possibly coupled with influence on longer (interannual) scales that we were not able to analyse due to the finite size of our time series \citep{Blesic2018}. The observed loss of seasonality poses a substantial risk for the stability and functioning of riverine ecosystems, particularly for systems strongly adapted to seasonal and interannual flow variability, raising, among other, economic and food security concerns \citep{Mor2018, Yang2018, HernandezGuzman2016, Sa-Oliveira2015, Benedick, Nikolic2018}. According to our results, this alteration in seasonality caused by damming extends beyond the reservoir, up to 220 km upstream, and affects river dynamics of Danube and of its tributaries.
	
	Additional cycles that appeared in our WT spectra, or were enhanced by the construction of dams - the 24-day cycle in Brza Palanka (MD) station, or a 15-day and even 40-day cycles in Prahovo (DS) and Golubac (UP) stations, can not be connected to the flow regulation in a straightforward manner, and could arise as results of either pure economic, that is, human-made influence, or climatic and/or hydrodynamic events that dam construction reinforces. On the Northern hemisphere, there exist two important modes of oscillation with periods near 48 and 23 days \citep{Ghil2002}. Both are dominated by zonal planetary wave number two. In addition, the 35-40 day oscillation, characterized by blocking structure over Eurasia continent was found by \citep{Zhang1997}. The main intra-seasonal oscillation in tropics, the Madden–Julian Oscillation (MJO), with quasi-regular period from 30 to 60 days \citep{Zhang2005} can also be linked to precipitation regimes over Europe, since it has been shown that it influences the North Atlantic Oscillation (NAO), which on the other hand has profound influence on European weather and climate \citep{Cassou2008}. For tropical and subtropical precipitation, quasi-biweekly oscillation (QBW) is one of the most important components \citep{Krishnamurti2012} together with MJO, but still there is no strong evidence that QBW is connected with extratropical regions, even if some relations were found \citep{Kikuchi2009}. Therefore, the link between 15-day cycle in precipitation over Danube sub-catchment and some intra-seasonal time scale atmospheric mode can be considered unclear.
	
	On the other hand, regarding the (SFRY) energy planning practices, cycles in the water levels may be expected at the intervals of 1, 7, and 365 days \citep{Jakovljevic1979}. The interval of 3 days may be the result of occurrences which start at Friday noon and end at Monday noon, or may come out of patterns of Romanian planning \citep{Jakovljevic1979}. The longer intervals, at 15, 24, and 40 days should not be completely excluded as human-made events. They might result from rain forecasts that have been used by dam operators to adjust the operation of dams. Further research in this direction may also show an intraday periodicity. Also, further studies may find changes in the planning methodology that occurred at the dissolving of SFRY in 1990s, and after 2000s, and/or due to wholesale electricity trade liberalization. Human-related changes described in such a way might show the path for better utilization of dams in the regional power system and at the EU wide power exchange market. Also, better utilization of dams may be obtained by including the climate change and flood control effects \citep{Eum2012}.
	
	The records of major importance to this paper, those from the vicinity of Djerdap/Iron Gates dams, were of the river level measurements, and we were not able to analyze them in parallel to the corresponding river flow data. Even if the relations linking these two variable suggest that their scaling statistics should be the same, which is the result we obtained (and thus corroborated) using the DFA method here for the records of stations surrounding the damming region, we found that the WTS functions of river level data are more sensitive to cyclic changes induced by damming. It remains for this finding to be further examined on a larger dataset from hydrological stations with both sets of records. 
	
	Finally, it remains for future studies to assess changes that we observed in multifractality of river level data, connected to the construction of dams, in a more systematic manner. Our WTMM results should particularly be considered in relation to our WTS results that point to possible stabilization of scaling regimes in the period after the construction of the second dam. Namely, the WTS functions there do not present with appearance of new or reappearance of prominent peaks from the period after the construction of the first dam, but rather behave as WTS functions of noisy series, shifted to the new scaling regime after the construction of the first dam. This result needs to be additionally systematically studied, in relation to climatic and hydrological mechanisms, and particularly concerning the working regimes of Djerdap/Iron Gates dams.   
	
	%
	%
	%
	%
	%
	%
	%
	%

	\acknowledgments
	We would like to thank Davide Zanchettin and Angelo Rubino for the valuable introductory discussions on the influence of changes in river morphology to the change in river flow dynamics that lead to the formulation of this paper's research question. We are also thankful to Andreja Martinoli for an insight into the foundations of the optimal energy planning in SFRY. This work received funding from the European Union’s Horizon 2020 Research and Innovation Programme under the Marie Sk{\l}odowska-Curie Grant Agreement 701785 (SB), and Serbian Ministry of Science, Education, and Technological Development grants no. 171015 (DjS and SB) and 176013 (VDj). The list of hydrological stations that are maintained by RHMZS is provided at URL \linebreak http://www.hidmet.gov.rs/eng/hidrologija/izvestajne/index.php; data recorded from 1991 are available in PDF format at URL \linebreak http://www.hidmet.gov.rs/ciril/hidrologija/povrsinske\_godisnjaci.php, while longer historical records are available for research purposes upon specified written request at office@hidmet.gov.rs. Support for the Twentieth Century Reanalysis Project dataset is provided by the U.S. Department of Energy, Office of Science Innovative and Novel Computational Impact on Theory and Experiment (DOE INCITE) program, and Office of Biological and Environmental Research (BER), and by the National Oceanic and Atmospheric Administration Climate Program Office. Wavelet software was provided by C. Torrence and G. Compo, and is available at URL: http://atoc.colorado.edu/research/wavelets/.

	
	%

	%

	
	

\end{document}